\documentclass{article}

\usepackage[]{graphicx}
\usepackage{amsmath}

\newcommand{\Sh}{Schr\"odinger{ }}

\newcommand{\bs}[1]{\boldsymbol{#1}}

\newcommand{\fues}[1]{\left(#1\right)}

\newcommand{\yav}[1]{\left[#1\right]}

\newcommand{\abs}[1]{\left\vert#1\right\vert}

\newcommand{\Eq}[1]{eq. (\ref{#1})}

\title{A model for the microwave assisted zero resistance  states}

\author{Manuel Torres$^1$, Alejandro Kunold$^2$\\
$^1$Instituto de F\'{\i}sica,\\
Universidad Nacional Aut\'onoma de M\'exico,\\
Apdo. Postal 20-364, M\'exico D.F. 01000, M\'exico,\\
e-mail: {\sf torres@fisica.unam.mx.}, Phone: 56\,22\,51\,44\\
$^2$Departamento de Ciencias B\'asicas,\\
Universidad Aut\'onoma Metropolitana Azcapotzalco,\\
Av. San Pablo 180, Col. Reynosa Tamaulipas,\\
Azcapotzalco, M\'exico D.F. 02200, M\'exico,\\
e-mail: {\sf akb@correo.azc.uam.mx}, Phone: +53\,18\,93\,81}

\begin{document}

\maketitle   

\begin{abstract}
  In this work we present a  model for the photoconductivity of a  two dimensional electron system (2DES) subjected  to a magnetic field. The model includes the microwave and Landau contributions in a non-perturbative  exact way, Impurity scattering   effects are treated perturbatively.  Based on this formalism, we provide a  Kubo-like   formula that takes into account the oscillatory Floquet structure of the problem. We discuss results related with the 
  recently  discovered zero-resistance states. 
\end{abstract}

\section{Introduction.}
 Recently, two experimental groups \cite{mani1,mani2,zudov1,zudov2},  reported the observation of a novel phenomenon: the existence  of  zero-resistance states  in an ultraclean  $GaAs/Al_xGa_{1-x} As$ sample  subjected to microwave radiation and moderate magnetic fields.      The   magnetoresistance  exhibits strong oscillations  with regions of zero resistance governed by the ratio $\omega /\omega_c$, where $\omega_c$ is the cyclotron frequency.  According  to  Zudov $etal.$,  the   oscillation amplitudes  reach  maxima  at $\omega/\omega_c =j$ and minima at  $\omega/\omega_c =j + 1/2$, for $j$ an integer. On the other hand Mani. $etal.$ 
 reported also  a periodic oscillatory behavior, but with maxima at $\omega/\omega_c =j - 1/4 $ and minima at $\omega/\omega_c =j +1/4$.

In spite of a large number of theoretical works,  a complete understanding has not yet been achieved.
A pioneer work put forward by  Ryzhii \cite{ry1,ry2} predicted the existence of negative-resistance 
  states. Durst and collaborators \cite{durst}  also found negative-resistance states in a  a diagrammatic calculation of the photoexcited electron scattered by a disorder  potential.    A possible connection between the calculated negative-resistance states and the observed vanishing resistance was put forward in reference \cite{andre}, noting that a general analysis of Maxwell equations shows that negative resistance  induces an instability  that  drives  the system  into a zero-resistance state. 
  In our model,  the Landau-Floquet  states act coherently with respect to the  oscillating field of  the impurities, that 
  in turn induces  transitions from levels below Fermi to levels above it. This formalism is complemented with a generalization of the Kubo formalism in order to correctly include the Floquet   structure of the problem.

\section{The model.}
We consider the motion of an  electron in two dimensions, subject to a uniform magnetic  field  $\mathbf{B}$  
perpendicular to the plane  and driven by  microwave radiation. In the long-wave limit the 
 dynamics is governed by the \Sh equation 
\begin{equation}\label{ecs1}
i \hbar \frac{\partial \Psi }{\partial t}= H \Psi  =  \left[  H_{\{B,\omega\}}  + V({\bs r} )  \right] \Psi  \, , 
\end{equation}
here $H_{\{B,\omega\}}$ is the Landau hamiltonian coupled to the radiation 
$H_{\{B,\omega\}} = \frac{1}{2m^*} {\bs \Pi}^2,$ $m^*$ is the  effective electron mass,  $ {\bs \Pi} = \mathbf{p}+e\mathbf{A}$, and  the vector potential  $\mathbf{A}$  includes  the external magnetic field and  radiation field (in the $\lambda \to \infty $ limit) contributions:
 $ \mathbf{A} = - \frac{1}{2} \bs r \times \bs B  +  Re \,\, \left[  \frac{\bs E}{\omega} \exp\{ -i \omega t \} \right]$.  
The  impurity scattering potential is decomposed in a Fourier expansion 
$ V(\bs r) = \sum_i \int d^2 q \, V( \bs q) \exp\{i {\bs q}  \cdot \left( {\bs r }- {\bs r}_i \right)   \},$ 
where $ {\bs r}_i$ is the position of the $i$th impurity, and  the explicit form of the potential   coefficient  $V( \bs q) $
for neutral impurity scattering  is $V( \bs q) = \frac{2 \pi^2 \hbar^2 V_0}{m^* {\mathcal E}_F } $, whereas for charged impurities  localized within  the doped layer of thickness $d$:  $ V( \vec  q)  = \frac{\pi\hbar^2 }{m^* } e^{-q d}  /\left( 1 + \frac{q}{q_{TF}} \right),$ with    $q_{TF} = e^2 m^* \left(2 \pi \epsilon_0 \epsilon_b \hbar^2 \right) $.

A three step procedure is enforced  in order to solve the problem posed by \Eq{ecs1}: 
(1) The Hamiltonian   $H_{\{B,\omega\}}$   can be exactly diagonalized by a transformation of the form 
$W^{\dagger} H_{\{B,\omega\}}  W = \omega_c \left( \frac{1}{2} + a_1^\dagger \, a_1\right) \equiv H_0 $,
with the  $W(t)$ operator     given by 
\begin{equation}\label{opw}
W(t)=  \exp\{i \eta_1 Q_1\}  \exp\{i \xi_1 P_1\} \exp\{i \eta_2 Q_2\}  \exp\{i \xi_2 P_2\}   \exp\{i  \int^t  {\mathcal L} dt^\prime \}
 , \end{equation}
where  the functions $\eta_i(t)$ and $\xi_i(t)$ represent the solutions to the classical equations of motion  and the 
$(Q_\mu, P_\mu)$ operators  are  the generators of the electric magnetic translation symmetries  \cite{Ashby1,Kunold1}.
(2) The transformation induced by $W$ is applied to the \Sh \Eq{ecs1}, transformed into 
$ i \hbar \frac{\partial \Psi^{(W)}  }{\partial t}  =  \left( H_0  + V_W (t)  \right) \Psi^{(W)},$ 
where  $ V_W  (t)  = W(t) V( {\bs r  }) W^{-1}(t)$ and $\Psi^{(W)}= W(t)  \Psi$. Note that the impurity potential acquires a time dependence  brought by   the  $W(t)$ transformation. (3) The problem is now solved in terms of an evolution operator $U(t)$, using the  interaction representation  and first order time dependence perturbation theory.  The solution to the original \Sh equation in \Eq{ecs1} has been achieved  by means of three successive  transformations
\begin{equation}\label{3trans}
\vert  \Psi_\mu (t) \rangle  = W^\dag  \,  \exp\{-i H_0 t\} \,  U(t - t_0) \,  \vert  \mu \rangle .
\end{equation}
The explicit expressions for the  matrix element of these  operators in the Landau-Floquet base  appear in detail 
in reference \cite{Torres3}.

\section{Kubo formula for Floquet states.}
The usual Kubo formula for the conductivity must be modified in order to include the Floquet dynamics.
In the presence of an additional $DC$ electric field  the complete Hamiltonian is 
 $H_T = H + V_{ext},$
where $H$ is the Hamiltonian in \Eq{ecs1} and $V_{ext} = \frac{1}{m} {\bs \Pi} \cdot {\bs A}_{ext}$ with  ${\bs A}_{ext}= \frac{{\bs E}_0}{\omega} \, sin \left( \Omega t\right) \, exp \left( - \eta \vert t \vert  \right).$ 
The static limit is obtained with $\Omega \to 0$, and $\eta $ represents the rate at which the perturbation is turned on and off. In order to calculate the expectation value of the current density, we need the density matrix  $\rho(t)$  which obeys  the von Neumann equation 
\begin{equation}\label{vn1}
i \hbar \frac{\partial \rho  }{\partial t}  =  \yav{H_T , \rho} =  \yav{H + V_{ext}  , \rho}.
 \end{equation}
We write to first order   $\rho = \rho_0 + \Delta \rho$, where the 
density matrix $\rho_0$  satisfies the equation 
$ i \hbar \frac{\partial \rho_0  }{\partial t}  =  \yav{H , \rho_0} $.
In agreement  with  \Eq{3trans}   the density matrix is transformed as $\tilde {\Delta \rho}  (t)   = U^\dag  \exp\{i H_0 t\}  W {\Delta \rho(t)}  W^\dag  \exp\{-i H_0 t\}   U $, and  obeys the  equation 
 $ i \hbar \frac{\partial   \tilde {\Delta \rho}  }{\partial t}  =  \yav{\tilde{V}_{ext} , \tilde{ \rho}_0 },  $
where $\tilde{V}_{ext}$ and $ \tilde{ \rho}_0$ are the    external potential and quasi-equilibrium density matrix 
transformed  in the same manner as  $\tilde {\Delta \rho}$. The transformed  
quasi-equilibrium density matrix is assumed to have the form  
$ \tilde{ \rho}_0 = \sum_\mu \vert \mu \rangle f(\epsilon_\mu) \langle \mu \vert ,$ where $f(\epsilon_\mu)$  is the usual Fermi function and $\epsilon_\mu$ the Landau-Floquet levels. The argument  behind this selection  is an adiabatic assumption 
that the original   Hamiltonian $H$ produces a quasi equilibrium state 
characterized by the  Landau-Floquet eigenvalues.
We then obtain the expectation value of $\tilde {\Delta \rho}  (t)$  in the Landau-Floquet base, from which the current density  is evaluated according to  $\langle  {\bs J} (t, {\bs r} ) \rangle  = Tr \left[   \tilde {\Delta \rho}  (t)   \tilde {\bs J} (t)  \right] $. The macroscopic  conductivity tensor that relates the average current density  to the averaged electric field, is obtained performing a space-time average, as well as an impurity 
average, assuming  non correlated impurities.
We take into account the lifetime $\tau = 2 \pi/\eta$ of the  quasiparticles induced by the weak scattering,  in the usual Born approximation.  We work out explicit expressions  for the longitudinal and Hall conductivities 
 (both for the dark and microwave induced contributions) \cite{Torres3}, and quote the result for the longitudinal photoconductivity 
 \begin{equation}\label{condLw}
\langle  {\bs \sigma}_L^\omega \rangle  =   \frac{\pi e^2 n_I  }{\hbar}  
\int d\epsilon    \sum_{\mu \nu}  \sum_l   A\left(\epsilon -\epsilon_\mu \right)B^{(l)}   \left(\epsilon ,\epsilon_\nu \right)    \int d^2 q   
 K(\bs q) \bigg{\vert} J_l\left( \vert \Delta \vert \right)  V( \bs q)  D_{\mu\nu} (\tilde{q}) \bigg{\vert}^2,
 \end{equation}
 where $  n_I$ is the  two dimensional impurity density,  and the 
 broadened spectral function is given as \\
$ A(\epsilon- \epsilon_\mu)  = \frac{\imath}{2 \pi} \left[G^+_\mu (\epsilon) - G^-_\mu (\epsilon) \right],$  where 
  $G^{\pm}_\mu (\epsilon) = 1/\left(\epsilon -\epsilon_\mu  \pm \imath \eta/2 \right) $ are the advanced and  retarded  Green's functions with finite lifetime $2 \eta^{-1}/\hbar$.  The auxiliary functions $K(\bs q ) $
  and $ B^{(l)}$  are defined according to 
\begin{equation} \label{funaux} 
K(\bs q ) = \omega_c^2 l_B^2 \, \, \frac{  q_x^2 \left[ \left( \epsilon_{\mu\nu} + \omega l  \right)^2  + \eta^2 \right] + q_y^2 \omega_c^2   - 2 \omega_c q_x q_y \eta}
{\vert \left(\epsilon_{\mu\nu} + \omega l  - \imath \eta  \right)^2 - \omega_c^2 \vert^2}, 
 \end{equation}
and 
 \begin{equation} \label{derspec} 
 B^{(l)} ( \epsilon,  \epsilon_\nu) =  \left[  \frac{d}{d\epsilon_0}  \left\{ \left[ f( \epsilon+ l \omega + \epsilon_0)-  f ( \epsilon)\right]  A( \epsilon - \epsilon_\nu + l \omega + \epsilon_0) \right\}
  \right]_{\epsilon_0= 0}. 
 \end{equation}
 In \Eq{condLw}  $J_l\left( \vert \Delta \vert \right)$ is the Bessel function with
$ \Delta =  \frac{\omega_c l_B^2 e E }{\omega \left( \omega^2 - \omega_c^2 + i \omega \Gamma \right)}
\left[ \omega \left(q_x e_x + q_y e_y  \right)  + i \omega_c \left(q_x e_y  -  q_y e_x  \right) \right]$, 
  $l_B=  \sqrt{\frac{\hbar}{eB}}$  being the magnetic length,  $\Gamma$ the electron radiative decay width and
$\bs \epsilon$ the polarization vector. Finally    $D^{\nu \mu}\fues{ \tilde{q}}= e^{-\frac{1}{2}\abs{  \tilde{q}}^2}  \tilde{q}^{\nu-\mu}\sqrt{\frac{\mu!}{\nu!}} L^{\nu-\mu}_{\mu}\fues{\abs{ \tilde{q}}^2},$ with  $L^{\mu}_{\mu}$ being the generalized Laguerre polynomial and $\tilde{q} = i l_B (q_x - i q_y)/ \sqrt{2}$.
\begin{figure}[htb]
\includegraphics[width=12cm, height=8cm]{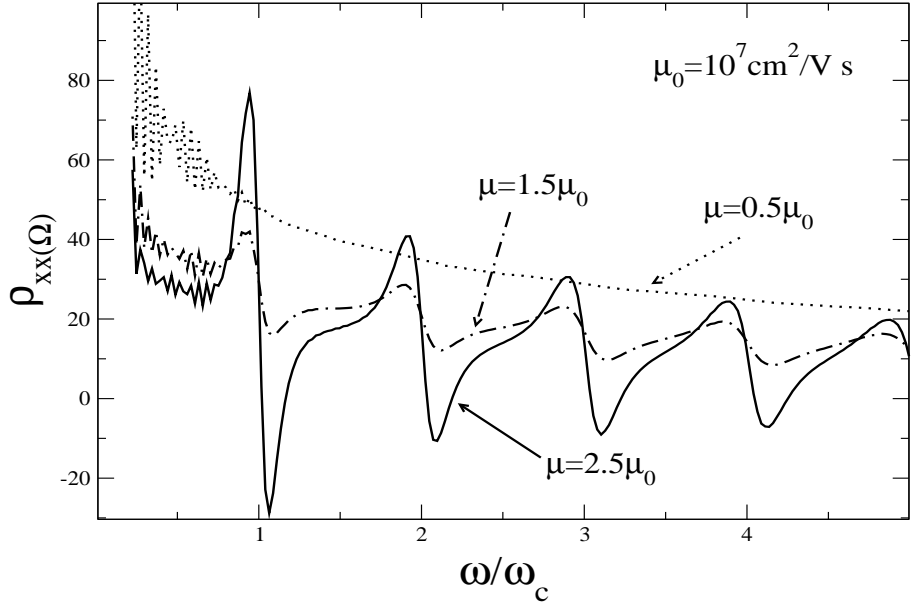}
\caption{Longitudinal resistance as a function of $\omega/\omega_c$ for three selections of the 
mobility.}
\label{figure1}
\end{figure}

\section{Results and conclusions} 

We have selected parameter values    corresponding to reported experiments  \cite{mani1,zudov1}:  effective  electron mass $m^* = 0.067 \, m_e$,     relative permittivity   $\epsilon_b \approx 13.18 $, fermi energy $\epsilon_F = 8 \, meV$, electron  mobility $ \mu \approx 0.5-2.5 \times 10^7 cm^2/V  s$, electron density $n = 3 \times 10^{11} cm^{-2}$,  microwave frequency $f  = 100 \, Ghz$,
magnetic fields in the range $0.05 - 0.4 \, \, Tesla$ and temperatures $T \approx 1 \,  K$.
Typical  microwave power is $10-40 \, nW$ from which  the   electric field intensity is estimated as  
$ \vert \vec E \vert  \approx 250 \, V/m$.  The  broadening $\eta$  is known to increase with the square root of the magnetic field,  hence we take  $\eta^2 =  \hbar \omega_c \left( 2 \pi \hbar  /\tau \right)$, which is connected through the relaxation time $\tau$ with the zero field  mobility  $\mu = e \tau /m^*$.  In the case of charged impurity scattering,  the distance  $d$  between the impurity and the 2DES  is  taken as   $ d \approx 30 \, nm$, and $n_I$ is estimated as $n_I \approx  7 \times 10^{11} \, cm^{-2}$. Explicit calculations demonstrate that 
results for  charged  neutral impurity  scattering are very similar \cite{Torres3}. Finally, the electron  radiative decay  $\Gamma$,  is interpreted as coherent dipole re-radiation of the oscillating 2D  electrons excited by microwaves;  hence, it is given by  $\Gamma = n e^2 /\left( 6 \epsilon_0 c \, m^* \right)$,  using the values of $n$ and $m^*$ given above, it yields  $ \Gamma \approx 0.38 \, meV $.

 One of the puzzling  properties of the observed giant magnetoresistance oscillations is
     related to the fact that they appear only in samples with an electron mobility exceeding  $  \approx 1.5 \times 10^7 cm^2/V  s$.  This phenomenon is absent  in samples in  which $\mu$ is reduced by   one-order of magnitude. This behavior is well reproduced   by the present formalism.  Figure (\ref{figure1})  displays the $\rho_{xx} \, \, vs. \,\, \omega / \omega_c $ plot for three  selected values of $\mu$.
   For $\mu \approx 0.5  \times 10^7 cm^2/V  s$ the expected almost linear behavior $\rho_{xx} \propto B$ is clearly depicted.  As the electron mobility  increases to $\mu \approx 1.5  \times 10^7 cm^2/V  s$ the magnetoresistance oscillations are clearly observed, but negative resistance states only  appear when  $\mu \approx 2.5  \times 10^7 cm^2/V  s$.
   We notice that $\rho_{xx}$ vanishes at $\omega /\omega_c =j$ for $j$ integer.  The oscillations follow a pattern with minima at $\omega/\omega_c =j + \epsilon $, and maxima at  
    $\omega/\omega_c =j -  \epsilon $, adjusted with $\epsilon \approx 1/10$. 
    
    Eqs.  (\ref{condLw}-\ref{derspec}) contain  the main ingredients that explain the huge increase observed in  the longitudinal conductance  when the material is irradiated by microwaves.  In the standard expression for the Kubo formula there are no Floquet  replica contribution, hence $\omega$ can be set to zero in (\ref{derspec}), in which case $B^{(l)}$ becomes proportional  to the energy derivative  of the Fermi distribution, that  in the $T \to 0$ limit  becomes of the form $\delta (\epsilon - \epsilon_F)$, and the conductivity depends only on  the states at the Fermi energy.  On the other hand as a result of the periodic structure induced by the  microwave radiation,  $B^{(l)}$ is  dominated by  terms proportional to 
   $   \frac{d}{d\epsilon}  A( \epsilon - \epsilon_\nu +  l \omega )  $, hence several  contributions arise  from  the transitions between different Landau levels. 

The model can also be used in order to test  quirality effects induced by the magnetic field.  In reference  \cite{Torres3}  we carried out calculations for different $E-$field polarizations with respect to the current. The  amplitudes of  the resistivity   oscillation  are  bigger for transverse  polarization as compared to longitudinal polarization. Selecting  negative circular polarization, the oscillation amplitudes get the maximum possible value.  Instead, the negative resistance states disappear for positive circular polarization. These results are easily understood, because for negative circular polarization and $ \omega \approx \omega_c$ the  electric field rotates in phase with respect to the  electron cyclotron rotation.  Finally, we mention that the    above discussed results are well described by single photon processes, however the present formalism is well suited   to explore the non-linear regime in which multiple photon exchange  play an essential role.

\section{Acknowledgements}
We acknowledge the financial support endowed by
CONACyT through grant No. \texttt{42026-F}, the Departamento de
Ciencias B\'asicas UAM-A through project No. \texttt{2230403} and
Rector\'{\i}a General de la UAM-A.

\end{document}